\documentstyle[multicol,aps,prl,epsf]{revtex}
\begin{document}

\draft

\title{
Gate-induced band ferromagnetism in an organic polymer 
}

\author{
Ryotaro Arita, Yuji Suwa$^1$, Kazuhiko Kuroki$^2$, and Hideo Aoki
}

\address{Department of Physics, University of Tokyo, Hongo,
Tokyo 113-0033, Japan}
\address{$^1$Advanced Research Laboratory, Hitachi Ltd., Higashi Koigakubo, 
Kokubunji, Tokyo 185-8601, Japan}
\address{$^2$Department of Applied Physics and Chemistry,
University of Electro-Communications, Chofu, Tokyo 182-8585, Japan}

\date{\today}

\maketitle

\begin{abstract}
We propose that a chain of five-membered rings 
(polyaminotriazole) should be ferromagnetic 
with an appropriate doping that is envisaged to be 
feasible with an FET structure.  The ferromagnetism 
is confirmed by a spin density functional 
calculation, which also shows that 
ferromagnetism survives the Peierls instability.  
We explain the magnetism in terms of Mielke and Tasaki's 
flat-band ferromagnetism with the Hubbard model. 
This opens a new possibility of {\it band ferromagnetism} 
in purely organic polymers.  
\end{abstract}

\medskip


\begin{multicols}{2}
\narrowtext
Since the discovery of conducting organic 
polymers\cite{Shirakawa74,Shirakawa77},
a great variety of studies have been performed
to pave a new way to further employ polymers or 
oligomers in realizing various functions, 
such as field-effect transistors (FETs)\cite{Garnier94} or 
electroluminescent diodes\cite{Friend99}. 
This is most recently highlighted  by a series of works by 
Batlogg and coworkers\cite{Schoen00,Schoen01}, 
where they have demonstrated that some molecular 
crystals (anthracene etc) and polymers (polythiophene) can 
not only be metallized, but even exhibit superconductivity. 

If a gate-induced superconducting plastic is possible,
a gate-induced ferromagnetic plastic is of no less interests.  
Ferromagnetism, usually a feature of d or f electron systems, 
taking place in $\pi$ electron systems 
has indeed been a theoretical and experimental challenge, for 
which an enormous amount of studies have been carried out. 
In particular, purely organic ferromagnets are interesting.  
A theoretical proposal was made by Shima and one of the 
present authors,\cite{Shima93} where graphites 
with superhoneycomb structures are considered,
but this has to do with Lieb's ferr{\it i}magnetism\cite{Lieb}, 
a kind of antiferromagnetism for unbalanced numbers of 
sublattice sites.  Organics exhibiting (single-)band 
ferromagnetism have yet to be synthesized.\cite{Allemand91}

Here we propose a novel possibility
for a {\it band ferromagnetism} in a {\it purely}
organic polymer that realizes another class of flat-band ferromagnetism 
due to Mielke and Tasaki mechanism.\cite{Tasaki-review}  
We first find that a chain of 
five-membered rings with a right choice of 
the functional group (polyaminotriazole) has 
a flat band.  When this band is made half-filled, 
where the doping is envisaged to be realized in an FET structure, 
we then show that the ground state is ferromagnetic with 
a spin density functional 
calculation, which also shows that 
the magnetism survives the Peierls instability.  
We finally confirm that this really has to do with 
the Mielke-Tasaki mechanism by mapping  
the $\pi$-orbital system to a tight-binding model 
with the Hubbard repulsion.  
This way we end up with a gate-induced ferromagnetic plastic. 

The problem of ferromagnetism in the Hubbard model, 
despite 
a long history since the 1960s\cite{Gutzwiller,Kanamori,Hubbard},
is still some way from a full understanding.  
One sensible way to study the problem is to take some 
characteristic situation where ferromagnetism may be 
rigorously proved.  
In fact, Mielke and independently Tasaki have proposed 
some classes of models where a flat band exists 
at the bottom\cite{Tasaki-review}, 
and have proved rigorously for the Hubbard model 
that the ground state is fully spin-polarized 
when the flat band is half-filled.  
When the flat band lies in between dispersive bands, 
the rigorous proof becomes inapplicable, but ferromagnetism 
is still expected 
when what is called the local connectivity condition for the basis vectors in 
the flat band is satisfied.  This amounts to 
a condition that adjacent ``Wannier" orbitals {\it have to} overlap, 
despite the flat dispersion, no matter 
how they are combined to minimize the orbit size, which is 
why spins tend to align due to Pauli's principle.  
The case of a middle flat-band is actually studied in the 
context of a model atomic quantum wire,\cite{Arita98} 
where the ferromagnetism is shown to be realized 
unless the repulsive interaction is below a critical value.  

Mielke-Tasaki flat bands are usually constructed from 
interferences between nearest-neighbor and second neighbor 
transfers, which requires special lattice structures 
especially in spatial dimensions greater than two. 
On the other hand, in quasi-one-dimensional chains 
we can conceive lattices having flat bands relatively easily\cite{Arita98}, 
since it is easier to make Wannier orbitals which
satisfy the connectivity condition in a 
chain than in a network.  
This is why we first look for some organic polymers.
Although we will have to consider weak three-dimensional couplings 
to estimate the actual T$_c$,\cite{3D} 
the possibility of band ferromagnetism in organic polymers
should be an intriguing avenue to explore.
Here we propose that one such system 
has a flat band.  We then show from both a band structure 
calculation and an exact diagonalization for the Hubbard model 
that the system should be ferromagnetic when the band is made 
half-filled with a gate-induced doping in an 
FET (field-effect transistor) structure recently 
developed by Batlogg {\it et al}.\cite{Schoen00,Schoen01}
 
\begin{figure}
\begin{center}
\leavevmode\epsfysize=15mm \epsfbox{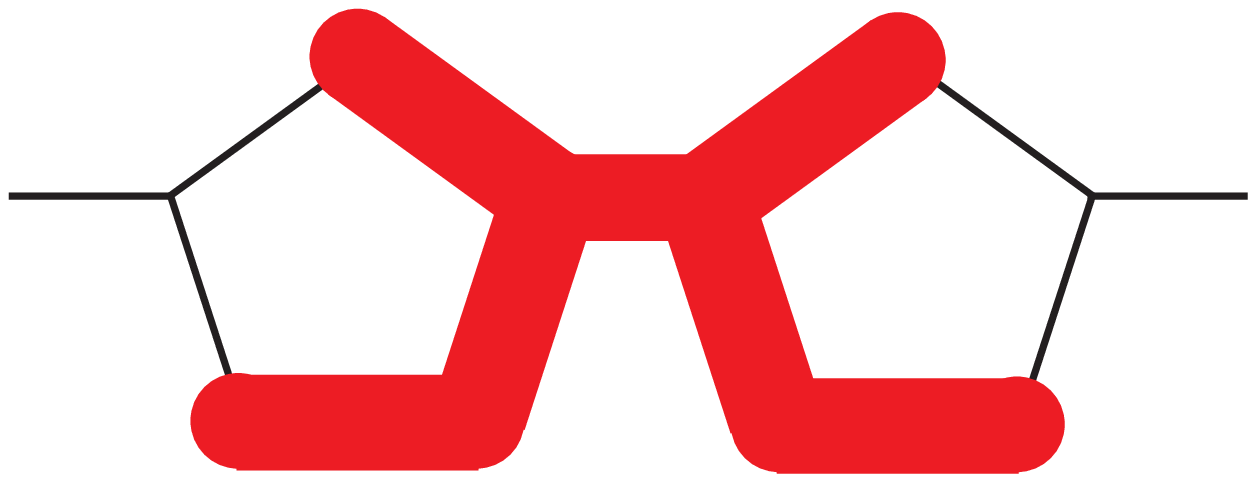}
\caption{The connected five-membered rings, where 
the shading indicates a ``Wannier" orbital 
that satisfies the local connectivity condition.}
\label{model}
\end{center}
\end{figure}

We first note that polymers comprising five-membered rings should be 
promising.  This is because we find it empirically easier to realize 
flat bands in chains of odd-membered rings.  
Intuitively, odd-membered rings incorporate frustrations for electron 
transfers, while even-membered (i.e., bipartite) 
rings have an obvious disadvantage of a tendency toward antiferromagnetism 
when the electron-electron repulsion is turned on, 
which should compete with ferromagnetism.  
The tight-binding model
on the connected five-membered rings
has indeed dispersionless bands in appropriate, realistic conditions, where  
the eigenstate on the flat band indeed  satisfies 
the local connectivity condition 
(i.e., overlapping orbitals, see Fig. \ref{model}).  

\begin{figure}
\begin{center}
\leavevmode\epsfysize=45mm \epsfbox{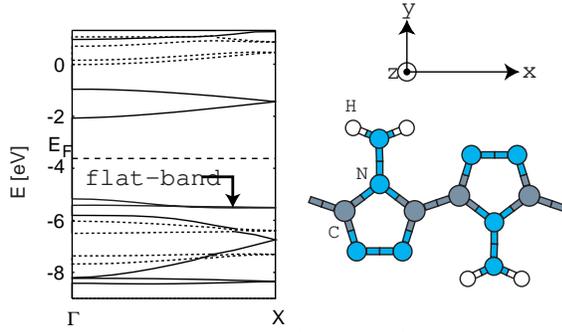}
\caption{The band structure (left panel)
and the optimized atomic configuration (right) 
for the (undoped) 
polyaminotriazole obtained by the GGA-DFT.
The solid (dotted) lines represent 
bands with $\pi$ ($\sigma$) character.
}
\label{ldaband}
\end{center}
\end{figure}

So we start with a search for the case of flat bands by scanning 
various five-membered polymers, i.e., 
polypyrrole, polythiophene, polytriazole, polyaminotriazole, {\it etc.}
The band structure is obtained with first principles calculations 
within the framework of 
the generalized gradient approximation 
based on the density functional theory 
(which we call GGA-DFT).\cite{Perdew1996}
We adopt the exchange-correlation functional introduced by
Perdew, Bruke and Wang\cite{Perdew1996} and
ultra-soft pseudo-potentials\cite{Vanderbilt90,Laasonen93} 
in a separable form. The wave functions 
are expanded by plane waves up to a cut-off energy of 20.25Ry.
As for the unit cell, five-membered rings usually alternate 
their directions in a chain (see Fig. \ref{ldaband}), 
so that there are two rings in a unit cell with the Brillouin zone folded.  
We take a repeated-chain model along directions perpendicular to the 
chain with a sufficiently large repeat distance.   
The atomic configuration as well as the unit cell size along the chain 
are optimized to minimize the ground state energy 
with the conjugate gradient scheme\cite{Yamauchi1996}.

It turns out that the flat band is rather hard to realize even for 
five-membered chains, which is not too surprising since 
an odd-membered ring is by no means a sufficient condition for 
a flat band.   However, attaching some functional group to each 
ring helps since this changes site energies and transfer integrals, 
and we have found that among the polymers investigated here
polyaminotriazole (poly(4-amino 1,2,4triazole) to be precise, 
see inset of Fig.\ref{ldaband}) hits the right condition.  
Figure \ref{ldaband} shows that the top valence band 
(with two branches due to the above-mentioned band folding) 
has little dispersion ($\sim$ O(0.1eV)) there.

\begin{figure}
\begin{center}
\leavevmode\epsfysize=45mm \epsfbox{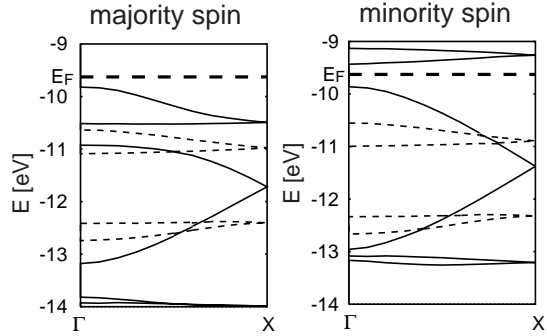}
\caption{The band structure of the {\it doped} system 
with an optimized structure 
obtained with the GGA-SDFT. The solid (dotted) lines represent
the bands with $\pi$ ($\sigma$) character.}
\label{lsdaband1}
\end{center}
\end{figure}

To induce the flat-band ferromagnetism, this band has to be 
half-filled, while the band is fully filled when undoped.  
To assess the possibility of the spin polarization 
in hole-doped polyaminotriazole, we have 
carried out the GGA including the spin degrees of freedom 
based on the 
spin density functional theory (which we call GGA-SDFT).\cite{Perdew1996} 
We have focused on the situation where the highest valence band
is hole doped to make its upper branch empty.
Figure \ref{lsdaband1} shows the 
band structure in which we have taken a polarized state as the
initial state. The optimized state remains to be polarized, 
where the splitting between the 
majority-spin and the minority-spin bands is $\sim 1$ eV.
This is the same order of the exchange splitting 
estimated in \cite{Okada00} for the atomic quantum wire.\cite{Arita98}

\begin{figure}
\begin{center}
\leavevmode\epsfysize=45mm \epsfbox{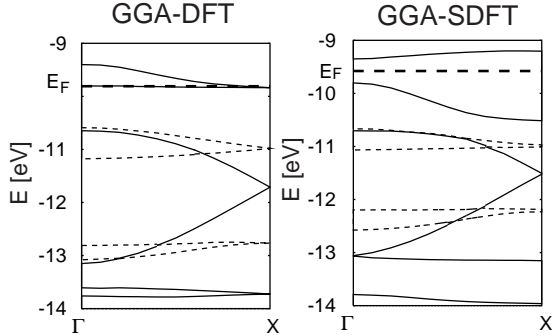}
\caption{Band structure for the doped system optimized by 
allowing the Peierls distortion in the GGA-DFT (left panel), 
and that for the antiferromagnetic state
obtained with the GGA-SDFT (right).}
\label{lsdaband2}
\end{center}
\end{figure}

We have to be careful when we dope the band, since one-dimensional 
metals are in general unstable against the 
Peierls distortion, where the electronic energy is lowered 
at the cost of the lattice distortion.  
So we have checked whether the energy gained due to
the spin polarization overcomes the Peierls transition.
The left panel of Fig. \ref{lsdaband2} 
is the band structure of the 
Peierls-distortion allowed, spin-unpolarized state obtained by 
the GGA-DFT. 
We can see that the Peierls-splitting 
at X is negligibly small.  Physically, this should be 
because $\sigma$-electrons which form the backbone 
of the structure have a rigid enough bonding 
that can cope with the Peierls instability here.   
The total energy of the global minimum in the
GGA-DFT calculation is higher than that of the polarized
optimized state in the GGA-SDFT calculation by $\sim$400 meV.

If we allow the spatial distributions of 
up and down spins to be different in the GGA-SDFT 
starting from an initially unpolarized state, 
the resulting band structure has
a wide gap at the Fermi level 
(right panel of Fig.\ref{lsdaband2}), 
which we attribute to an antiferromagnetic gap.  
Although the total energy of the antiferromagnetic state
is lower than that for the uniformly zero spin density 
(the left panel of Fig. \ref{lsdaband2}), 
the energy is higher than that of 
the optimized polarized state (Fig.\ref{lsdaband1}) by $\sim$50 meV.
Thus we can  conclude that 
the true ground state in the GGA-SDFT is the polarized state, 
so that we can expect that 
polyaminotriazole becomes, when appropriately doped, 
ferromagnetic at sufficiently low temperature.

\begin{figure}
\begin{center}
\leavevmode\epsfysize=45mm \epsfbox{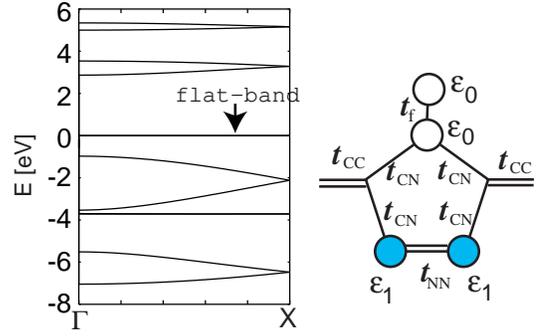}
\caption{The energy band dispersion of the tight-binding model
with $\epsilon_0=-1.43$ eV, $\epsilon_1=-0.5$ eV,
$t_{{\rm CN}}= t_{{\rm CC}} = t_{{\rm f}}=2.5$ eV, 
and $t_{\rm NN}=3.0$ eV.
To facilitate comparison with Fig. \ref{ldaband}, we have folded
the band to have a two-ring unit cell.}
\label{tightband}
\end{center}
\end{figure}

\begin{figure}
\begin{center}
\leavevmode\epsfysize=55mm \epsfbox{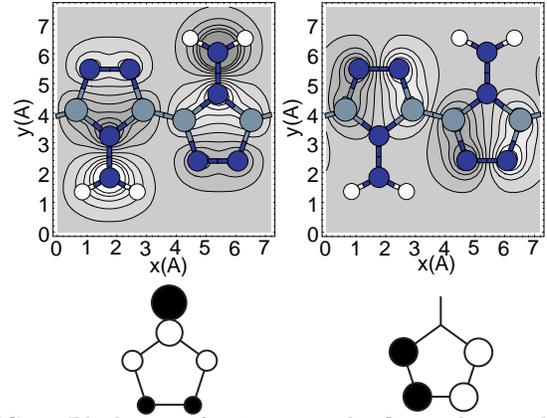}
\caption{Bloch wave functions on the flat $\pi$-electron bands 
in polyaminotriazole obtained with the GGA-DFT (top panels) are 
compared with 
the corresponding ones in the tight-binding model
(bottom panels).  Left/right panels correspond to 
two eigenstates on the flat band at $\Gamma$.
White/black regions (or circles)  
represent the sign of the wavefunction, while 
the size of the circles the amplitude.}
\label{wavefn}
\end{center}
\end{figure}

Now we come to the key question of whether the ferromagnetism 
obtained in the band calculation can be identified as the 
flat-band ferromagnetism for the Hubbard model 
{\it a l{\'a}} Mielke-Tasaki.   To do so we have first to map 
the $\pi$-electron system to a tight-binding model 
(see the right panel of Fig. \ref{tightband}), with a Hamiltonian,
${\cal H}_0 = -\sum_{ij\sigma}t_{ij} c^\dagger_{i\sigma}c_{j\sigma} 
+\sum_{i}\epsilon_i n_i$ in the standard notation.
We have estimated the values of transfer integrals 
by first identifying $\pi$ bands, then 
characterizing the wave functions as bonding or 
anti-bonding at Brillouin zone center and 
edges.  The site energies are estimated from the energy levels
of isolated atoms.  The obtained figures are: 
$t_{{\rm CN}}\simeq t_{{\rm CC}} \simeq t_{{\rm f}} = 2.5$ eV, 
$t_{{\rm NN}}= 3 \sim 4$ eV, 
$\epsilon_0 \simeq \epsilon_1 = -1 \sim -2$ eV.  
The band structure for this tight-binding model, 
depicted in the left panel of Fig.\ref{tightband}, 
indeed reproduces the features of that for 
$\pi$-electrons in polyaminotriazole (Fig. \ref{model}).
If we turn to wave functions in Fig. \ref{wavefn}, 
the wave functions at $\Gamma$
on the flat band in the tight-binding model also 
reproduce the features of  
those on the flat band in the GGA-DFT result for 
polyaminotriazole.  So we can conclude that the tight-binding 
model captures the band structure of polyaminotriazole.

\begin{figure}
\begin{center}
\leavevmode\epsfysize=50mm \epsfbox{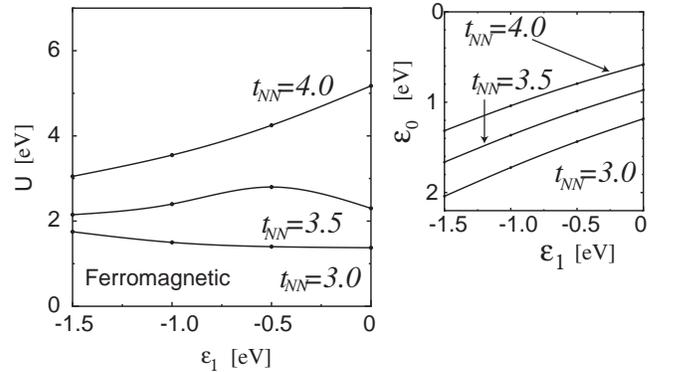}
\caption{Phase diagram for the Hubbard model 
on a lattice depicted in Fig. 5 against $U$ and $\epsilon_1$ 
for various values of $t_{\rm NN}$ for $t_0=2.5$ eV. 
The inset shows the relation between $\epsilon_1$
and $\epsilon_0$ to realize the flat band.}
\label{phase}
\end{center}
\end{figure}

We can then proceed to the question of 
whether the ground state is spin-polarized 
in the presence of the Hubbard interaction, 
$
{\cal H}_{U}=
U \sum_{i} n_{i\uparrow}n_{i\downarrow}.
$
We can draw a phase diagram by noting that 
a flat band arises when a set of equations, 
\begin{eqnarray*}
\varepsilon-\epsilon_1&=&(1-\varepsilon)
(t_{\rm NN}^2-(\varepsilon-\epsilon_1)^2)-t_{\rm NN}, \\
(\varepsilon-\epsilon_1+t_{\rm NN})/(1-\varepsilon)
&=&-t_{\rm NN}/(\epsilon_0-\varepsilon)
-t_{\rm NN}(\varepsilon-\epsilon_0),
\end{eqnarray*}
are satisfied, where $\varepsilon$ is the eigenenergy of the flat band 
and $t_{{\rm CN}}= t_{{\rm CC}} =t_0 (=1$ here) is assumed 
for simplicity.  
In Fig. \ref{phase} 
we show the phase diagram against $U$ and $\epsilon_1$ 
obtained with an exact diagonalizaion calculation for a 12-site 
(2 unit cell) Hubbard model for $t_0=2.5$ eV
and various values of $t_{\rm NN} = 3.0 - 4.0$ eV, 
where $\epsilon_0$ is chosen to satisfy the above equations 
throughout. 
We can see that we have indeed a ferromagnetic phase 
unless the repulsion is not too strong (i.e., $U < U_c$ 
with $U_c = 2 \sim 5$ eV).  The presence of a $U_c$ is 
as expected from the above discussion for a flat band 
lying in between dispersive ones.  
Our preliminary quantum Monte Carlo calculation suggests that
two unit cell is sufficient to roughly determine 
the ferromagnetic phase boundary.

Let us finally comment on the robustness of the 
flat-band ferromagnetism.  First, the one-electron 
dispersion does not have to be exactly flat to 
realize ferromagnetism, as confirmed from a number of 
studies.\cite{Kusakabe,Penc}  
Also, electron-electron interactions extending 
beyond  the on-site do not necessarily degrade 
the ferromagnetism.    
This has been shown for the extended Hubbard model 
where the off-site 
repulsion $V$ does not degrade, or in some situation 
even induces, the ferromagnetism\cite{Shimoi}.
As for the doping dependence,
a numerical calculation\cite{Sakamoto} 
has indicated that the ferromagnetism 
survives when the flat band 
is shifted away from the half-filling.  
Therefore, we may expect that the appropriate polymers 
as exemplified by polyaminotriazole should have 
ferromagnetic instabilities at low temperatures when 
the system is sufficiently doped in e.g. an FET geometry.  
In view of a quite recent finding\cite{Schoen01} that 
polythiophene may be doped 
with this geometry to realize superconductivity, 
we expect the gate-induced band ferromagnetism 
in purely organic polymers conceived here should be 
within experimental feasibility.  

We would like to thank A. Koma, K. Saiki, T. Shimada, K. Ueno,
Y. Shimoi, T. Hashizume and M. Ichimura for fruitful discussions.
This work was supported in part by a Grant-in-Aid for Scientific 
Research and Special Coordination Funds  
from Ministry of Education, Culture, Sports, Science and 
Technology of Japan.
The GGA calculation was performed with TAPP (Tokyo Ab-inito 
Program Package), where RA and YS would like to thank 
J. Yamauchi for technical advices.
Numerical calculations were performed on 
SR8000 in ISSP, University of Tokyo.

\end{multicols}

\begin{references}
\bibitem{3D} Ferromagnetism will remain in a bunch of chains if 
the magnetic easy axis is normal to the chain.
\end{references}
\end{document}